# Title: Self-Assembled Periodic Nanostructures Using Martensitic Phase Transformations

**Authors:** Abhinav Prakash[1,*,†], Tianqi Wang[1,†], Ashley Bucsek[2,3,†], Tristan K. Truttmann[1], Alireza Fali[4], Michele Cotrufo[5], Hwanhui Yun[1], Jong-Woo Kim[6], Philip J. Ryan[6], K. Andre Mkhoyan[1], Andrea Alù[5,7], Yohannes Abate[4], Richard D. James[2], and Bharat Jalan[1,*]

**Affiliations:**

[1]Department of Chemical Engineering and Materials Science, University of Minnesota – Twin Cities, Minneapolis, MN 55455, USA.

[2]Department of Aerospace Engineering and Mechanics, University of Minnesota – Twin Cities, Minneapolis, MN 55455, USA.

[3]Department of Mechanical Engineering, University of Michigan, Ann Arbor, MI 48109, USA.

[4] Department of Physics and Astronomy, University of Georgia, Athens, GA 30602, USA.

[5]Photonics Initiative, Advanced Science Research Center, City University of New York, NY 10031, USA.

[6]Advanced Photon Source, Argonne National Laboratory, Argonne, IL 60439, USA.

[7]Physics Program, Graduate Center, City University of New York, New York, NY 10016, USA.

*Correspondence to: Abhinav Prakash (praka019@umn.edu), Bharat Jalan (bjalan@umn.edu).

†Equal contributions.

**Abstract:** We describe a novel approach for the rational design and synthesis of self-assembled periodic nanostructures using martensitic phase transformations. We demonstrate this approach in a thin film of perovskite $SrSnO_3$ with reconfigurable periodic nanostructures consisting of regularly spaced regions of sharply contrasted dielectric properties. The films can be designed to have different periodicities and relative phase fractions via chemical doping or strain engineering. The dielectric contrast within a single film can be tuned using temperature and laser wavelength, effectively creating a variable photonic crystal. Our results show the realistic possibility of designing large-area self-assembled periodic structures using martensitic phase transformations with the potential of implementing "built-to-order" nanostructures for tailored optoelectronic functionalities.

**One Sentence Summary:** Martensitic transformations are used for the rational design of a perovskite metamaterial with reconfigurable high-contrast dielectric properties.

**Main Text:** Periodic nanostructures have enabled multiple applications pertaining to their tailored mechanical, electrical, magnetic and optical properties. For instance, periodic nanostructures with refractive index contrast at optical wavelengths have opened unique opportunities in optoelectronics, nanophotonics, and quantum optics, culminated with the surge of interest in metamaterial technology.[1-3] Fabrication of such structures, however, typically requires multi-step lithography processes with appreciable complexity, time requirements and cost. We show a way to create periodic nanostructures with large optical contrast in a self-assembled process over large areas by exploiting martensitic phase transitions in phase change materials. Martensitic phase transformations are diffusionless transformations between a high-symmetry and a low-symmetry crystallographic phase. Each phase can have distinct mechanical, electrical, magnetic, and/or optical properties, which is why this ideally reversible transformation gives way to the functional abilities of many ferroelastic, ferroelectric, and ferromagnetic materials[4-6]. When there is a thermal hysteresis, the two phases can stably coexist over a defined temperature range and form a microstructure and, within this microstructure, the two phases can only meet at specific interfaces (like puzzle pieces that can only fit in a specific way). This finite set of interfaces can be calculated using the rules of kinematic compatibility between the phases' point group symmetries and lattice parameters[7,8]. The less disparate the degrees of symmetry between the two phases are, the fewer possible interfaces exist[9,10]. For this reason, martensitic phase transformations can be used to engineer microstructures and nanostructures that consist of phases segregated into specific, predictable large-area patterns where each phase may have distinct properties.

The crystallographic theory of martensite (CTM) is a framework that can be used to predict all of the kinematically compatible interfaces between two phases that are related through a martensitic phase transformation[7,8]. Commonly, the two phases are not initially compatible with each other, and the low-symmetry phase must form a fine periodic substructure (i.e., twins) to improve the compatibility with the high-symmetry phase[11]. However, in thin films, the out-of-plane direction can be considered free, reducing the constraints to two dimensions and relaxing the requirements of kinematic compatibility.[9,12] For this reason, it is much easier for two phases to form kinematically compatible phase interfaces (without the need for twins) in a thin film. For example, when a tetragonal-to-orthorhombic martensitic phase transformation occurs in a thin film, there are two compatible interfaces between the phases where both interfaces are perpendicular to the substrate. This case is shown schematically in Figure 1(a). These two interfaces can be predicted from CTM using the two phases' point groups and the ratios of the tetragonal lattice parameters to the orthorhombic lattice parameters, $\alpha$, $\beta$, and $\gamma$.[11] Because there are only two possible interfaces in this case, it is useful to define the angle between them as $\omega$ (identified in Figure 1(a)), which again depends on $\alpha$, $\beta$, and $\gamma$. See supplementary text for details.

Because of the difference between the tetragonal and orthorhombic lattice parameters, a partial phase transformation can provide strain relief to a thin epitaxial film. This idea is akin to the well-known Clausius-Clapeyron relationship and strain engineering techniques commonly used in epitaxial films.[13] Furthermore, because the epitaxial strain is uniformly distributed across the film, this partial phase transformation occurs across the entire film, forming a structural periodicity as shown in Figure 1(a). The fineness of the phase domains results from a balance between the interfacial energy increase and elastic strain energy decrease associated with each interface.[11,12] Figure 1(b) shows $\omega$ as a function of local strain relief produced by the tetragonal-to-orthorhombic transformation in the two in-plane directions, $\varepsilon_{11}$ and $\varepsilon_{22}$. As the sample is heated, the epitaxial strain is expected to vary slightly due to thermal expansion. As a result, the effective tetragonal and orthorhombic lattice parameters slightly vary, causing $\alpha$, $\beta$, and $\gamma$ to vary, and thus causing $\omega$ to change.

Guided by the theory of tetragonal-to-orthorhombic martensitic phase transformations, thin films of La-doped $SrSnO_3$ (LSSO) with an undoped $SrSnO_3$ (SSO) buffer layer were grown on a $GdScO_3$ (GSO) (110) substrate using hybrid molecular beam epitaxy.[14,15] Figure 2 shows that the co-existence of the tetragonal and orthorhombic phases occurs more prevalently above a critical thickness. Figure 2(a–c) shows wide-angle x-ray diffraction (XRD) patterns for three films grown with different thicknesses. For all three patterns, the XRD peak at $2\theta \approx 44°$ corresponds to the pseudocubic (002) peak of the SSO tetragonal phase.[14] The thinnest sample (Figure 2(a)) appears to be predominantly tetragonal, while the two thicker samples (Figure 2(b–c)) show a second XRD peak at a slightly higher $2\theta$ value, corresponding to the SSO orthorhombic phase.[14] Figures 2(d–f) show nanoscale scattering type scanning near-field optical (s-SNOM) 2nd harmonic amplitude micrographs of the films from figures 2 (a–c), respectively, where the relatively dark (red/orange) and bright (yellow) contrast indicates starkly different regions of dissimilar dielectric properties. Based on the x-ray diffraction scans, it can be deduced that the darker regions correspond to the tetragonal phase, and the brighter regions correspond to the orthorhombic phase. Additionally, nanoscale periodicity was confirmed by the temperature-dependent rocking curve showing a periodic modulation in the intensity across the center film peak suggesting periodic microstructure consistent with the results of SNOM (SI, Figure S1). Figure 2(g) shows atomic-resolution annular dark-filed (ADF) scanning transmission electron microscopy (STEM) image of one of the phase interfaces in a 30 nm LSSO / 108 nm SSO / GSO (110) sample. The phases were identified by the presence of characteristics spots in image fast Fourier transforms (FFTs) shown in the insets. The FFT of the orthorhombic phase showed additional diffraction peaks. The interface is marked with a white dotted line for clarity. The surface roughness of such thin films was further studied using atomic force microscopy (AFM) attesting to a surface with low root mean square roughness (SI, Figures S2, S3). Sharp interfaces and an atomically smooth surface suggest that photonic devices based on an architecture that involves a one-step fabrication such as molecular beam epitaxy can potentially have low attenuation or transmission loss. As predicted by the CTM for tetragonal-to-orthorhombic martensitic phase transformations in thin films, there are only two types of phase interfaces present (an approximately vertical interface and an approximately horizontal interface in the s-SNOM micrographs), resulting in a periodic or quasiperiodic nanostructure. Because the two phases have different optical properties as indicated by the s-SNOM amplitude images (see Figure S4 of SI for comparison of s-SNOM contrasts and topographical changes), these periodic nanostructures are akin to 2D photonic crystals and are suitable for nanophotonic applications.[3]

As an example, Figure 3a shows a metagrating[16], obtained by alternating the two phases assuming a refractive index contrast of 3:1 (similar to the one measured in our structure shown in Figure 4(f)), placed at a distance $h$ from a mirror. By making the unit cell of the metagrating asymmetric and properly tailoring its geometry we can obtain exotic optical responses, such as wavefront steering with large efficiency[16], whereby a normally incident beam is reflected into only one of the three available diffraction orders with almost unitary efficiency (Fig. 3b). The periodicity of the metasurface determines a discrete number of diffraction orders, and the asymmetry of the unit cell enables suppressing coupling to the unwanted orders (Fig. 3a), so that all the reflected energy is steered towards the angle of choice (Fig. 3b). In order to enable this functionality over an ultrathin metasurface, large area and strong optical contrast are required[17], typically achieved with slow and costly processes such as e-beam lithography. The proposed self-assembly technique enables a fast, cheap and large-area process to fabricate similar structures, which provides an interesting pathway for metasurface applications.

To study the origin and evolution of the tetragonal–orthorhombic microstructure, a set of films were synthesized as a function of SSO buffer layer thickness and La concentration (Figure 4). Figure 4(b) shows the evolution of the tetragonal and orthorhombic phases as a function of SSO buffer layer thickness holding La concentration constant. At very small SSO buffer layer thickness (~ 2 nm), the film is predominantly tetragonal. As the SSO buffer layer thickness increases, the relative volume of the orthorhombic phase increases, forming a tetragonal–orthorhombic phase mixture as shown in Figure 2(e–f). Eventually, the film becomes predominantly orthorhombic. Note that these changes occur without any shift in the film peak positions, meaning the out-of-plane lattice parameter for the tetragonal and orthorhombic phases remains constant for all films regardless of layer thickness. This is indicative of the coherent strain state of both the phases with the underlying substrate and was confirmed using reciprocal space maps (SI, Fig. S5).

These measurements show that the partial tetragonal-to-orthorhombic martensitic phase transformation is a source of strain relief. The theoretical critical thickness for strain relaxation via the formation of misfit dislocations for a SSO film grown on GSO is ~10 nm, calculated using the Matthew-Blakeslee equation.[18] The much larger thicknesses that are observed here can be attributed to the co-existence of the two phases. That is, the release of the volumetric strain energy happens via the formation of the orthorhombic phase rather than forming misfit dislocations. The film grows initially in the tetragonal phase and, with increasing thickness, partially transforms to the orthorhombic phase via martensitic phase transformation. Eventually, at a thickness much larger than the theoretical critical thickness for dislocation formation, the film becomes thick enough to form dislocations. The film with a $t_{buffer}$ = 216 nm showed the signature of strain relaxation via the formation of misfit dislocations: The XRD peak shifted towards a higher $2\theta$ value corresponding to a lower lattice parameter.

A similar evolution of the tetragonal–orthorhombic microstructure is seen when the La dopant concentration in the LSSO layer is reduced. Figure 4(d) shows the evolution of the tetragonal and orthorhombic phases as a function of La cell temperature holding SSO buffer layer thickness constant at 8 nm, where an increasing La cell temperature corresponds to a higher La concentration. As the La concentration increases, the film goes from a tetragonal–orthorhombic to predominantly tetragonal suggesting dopant as a tunable parameter for the microstructure. Figure 4(e–f) shows infrared s-SNOM $2^{nd}$ harmonic amplitude micrographs where the microstructure is shown in terms of dielectric contrast for the films grown with La cell temperatures of 1150°C and 1200°C, respectively. With increasing La content (i.e., higher La cell temperature), the density of the orthorhombic phase (bright yellow) decreases, consistent with XRD results in Figure 4(d) (also see Fig. S6). These observations collectively show that the relative volume of the tetragonal and orthorhombic phases can be tuned using strain (controlled by either layer thickness or substrate selection) and defects (controlled by doping) resulting in different overall dielectric behaviors in such photonic crystals.

Besides the dielectric tunability of different films using strain and defects, the microstructure and resultant dielectric behaviors can be tuned within the same film using external stimuli such as temperature and the wavelength of an excitation laser. Figure 5 shows s-SNOM amplitude micrographs of the same film taken at different temperatures and wavelengths, where the microstructure is shown in terms of dielectric contrast. When the temperature is increased and the wavelength is fixed (Figure 4(a–c)), the angle between the two-phase interfaces ($\omega$ in the schematic shown in Figure 1(a)) becomes more acute while the relative volume of the two phases remains constant. The change in $\omega$ is consistent with our theoretical prediction as illustrated in Figure 1 and is attributed to the small change in the lattice parameters with temperature. On the other hand, when the wavelength is increased and the temperature is fixed (Figure 5(d–f)), the relative volume of the orthorhombic phase increases while the angle between the two-phase interfaces remains the same. These results show that the photonic crystal microstructure can be tuned for a specific application using external stimuli.

We have shown that martensitic phase transformations can be used for the rational design and synthesis of self-assembled large-area periodic metamaterials. The microstructure of these materials follows strict rules of kinematic compatibility and, as a result, can be predicted and optimized using theory, such as CTM. When the two phases involved in a martensitic phase transformation have sharply contrasting mechanical, electrical, magnetic, and/or optical properties, this design approach can be used to create metamaterials with tunable, predictable mechanical, electrical, magnetic, and/or optical behaviors. We have demonstrated this approach on SSO, which undergoes a tetragonal-to-orthorhombic martensitic phase transformation. Due to the relatively small symmetry disparity between the tetragonal and orthorhombic phases, when these two phases coexist in a thin film, there are only two possible compatible phase interfaces. The result is a periodic checkerboard-like microstructure where the periodicity and relative phase fractions can be tuned using strain engineering or doping. In SSO, the tetragonal and orthorhombic phases also have contrasted optical properties at different wavelengths, making this metamaterial appropriate for 2D nanophotonic crystal applications. Furthermore, the optical contrast within a single film can be tuned using temperature and laser wavelength, creating a sort of variable photonic crystal. These results demonstrate the unique advantages of using martensitic phase transformations to engineer self-assembled nanostructures in thin epitaxial films.

**Acknowledgments:**

**Funding:** This work was primarily supported by the Air Force Office of Scientific Research (AFOSR) through Grant No. FA9550-19-1-0245 and partially through NSF DMR-1741801 and the UMN MRSEC program under award no. DMR-1420013 and the AFOSR MURI through Grant No. FA9550-17-1-0002. The work also benefitted from the RDF Fund of the Institute on the Environment (UMN), the Norwegian Centennial Chair Program (NOCC), and two Vannevar Bush Faculty Fellowships. Work at the UMN involving thin film characterization using synchrotron x-rays was supported by the U.S. Department of Energy through DE-SC0020211. Parts of this work were carried out at the Minnesota Nano Center and Characterization Facility, University of Minnesota, which receives partial support from NSF through the MRSEC program. Use of the Advanced Photon Source, an Office of Science User Facility operated for the U.S. Department of Energy (DOE) Office of Science by Argonne National Laboratory, was supported by the U.S. DOE under contract no. DE-AC02- 06CH11357. A.F. acknowledges support from the NSF Grant No. 1553251 and Y.A. acknowledge support from AFOSR Grant No. FA9559-16-1-0172.

**Author contributions:** A.P., T.Q., and B.J. conceived the idea and designed the experiments. A.P., T.Q. and T. T. grew and characterized samples. A.B. and R.J. performed the CTM model calculations. A.F and Y.A. carried out SNOM measurements. M.C. and A.A. performed simulations. H.Y and A.M performed TEM studies. J.-W.K and P.J.R. performed synchrotron x-ray measurements and analyzed the data. A.P. A.B. and B.J. wrote the manuscript. All authors contributed to the discussion and manuscript preparation.

**Competing interests:** The authors declare no competing interests.

**Data and materials availability:** All data needed to evaluate the conclusions of the paper are present in the paper and/or the Supplementary Materials. Additional data related to this paper may be requested from the authors.

**Supplementary Materials:**

Materials and Methods

Figures S1-S6

References

**Figures (color online)**

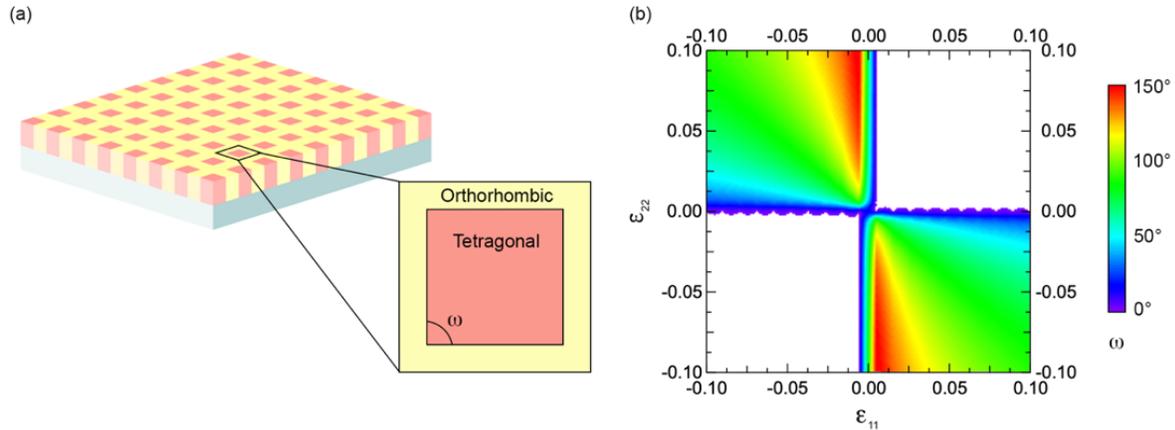

**Figure 1:** (a) Schematic of a self-assembled periodic nanostructure in a thin epitaxial film composed of a tetragonal phase (red) and an orthorhombic phase (yellow) related through a tetragonal-to-orthorhombic martensitic phase transformation. The schematic shows only two possible tetragonal–orthorhombic interfaces that are perpendicular to the substrate and are related by an angle ω. (b) A color map showing the angle between the two possible tetragonal–orthorhombic interfaces, ω, as a function of the local strain relief the tetragonal-to-orthorhombic martensitic phase transformation would produce in the two in-plane directions.

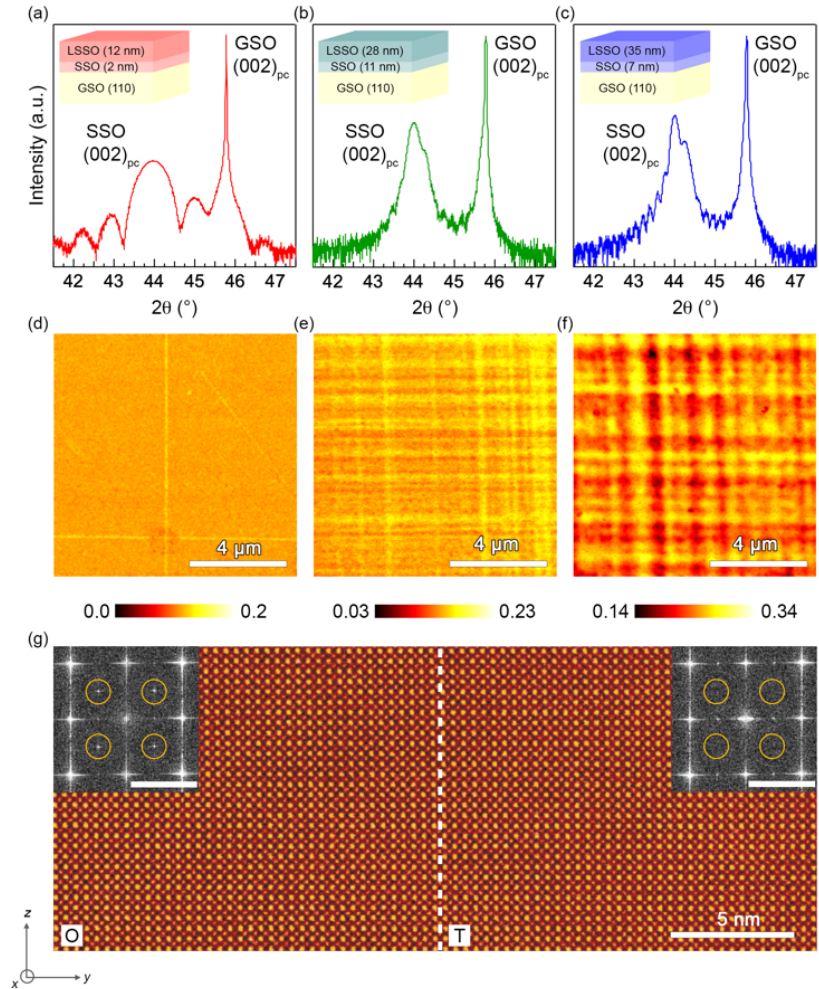

**Figure 2**: X-ray diffraction patterns of (a) 12 nm LSSO / 2 nm SSO / GSO (110), (b) 28 nm LSSO / 11 nm SSO / GSO (110), and (c) 35 nm LSSO / 7 nm SSO / GSO (110) films showing purely tetragonal phase in (a) and a mixture of tetragonal and orthorhombic phases in (b) and (c); (d), (e), and (f) are the corresponding s-SNOM images (with *false color*) revealing the dielectric character of the microstructure in these films; (g) cross-sectional HR-STEM (*false color*) of the interface between the tetragonal (T) and orthorhombic (O) phases in 30 nm LSSO / 108 nm SSO / GSO (110), where the insets show selected area Fourier transforms. Scale of the insets is 3 nm$^{-1}$.

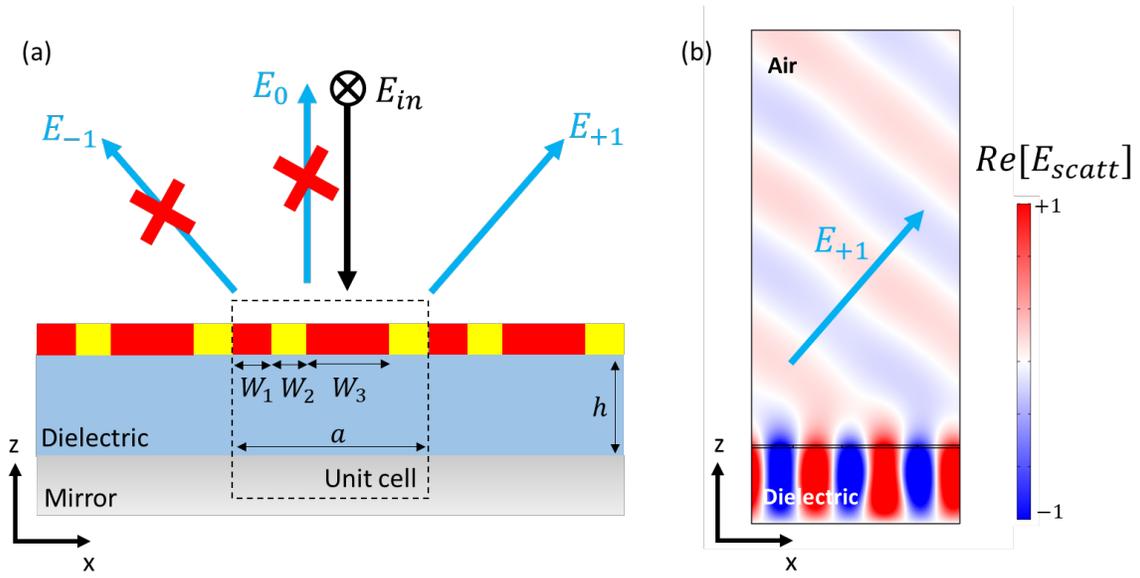

**Figure 3:** Example of nanophotonic application achievable by controlling the lateral widths of the two phases and their optical contrast. (a) A metagrating is designed such that a normally impinging beam is steered into only one of the three available diffraction orders, with efficiency > 98%. The structure is uniform along the $y$ direction. (b) Full-wave simulations of the metagrating response. The color plot shows the scattered field. Impinging wavelength $\lambda = 5\ \mu m$. Geometry: $a = 8\ \mu m$, $W_1 = 0.199a$, $W_2 = 0.235a$, $W_3 = 0.363a$, $h = 2.91\ \mu m$.

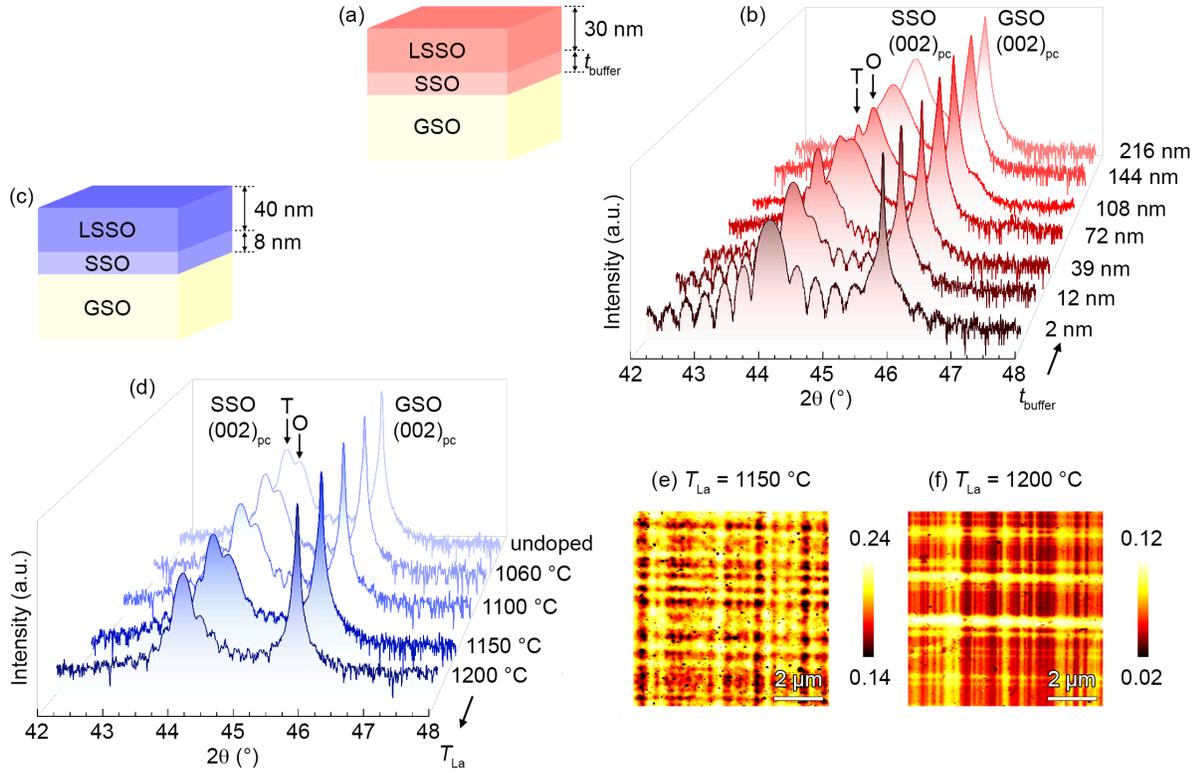

**Figure 4**: (a) Schematic of 30 nm LSSO / $t_{buffer}$ SSO / GSO (110) films with different SSO buffer layer thickness and fixed La concentration and (b) corresponding X-ray diffraction patterns. (c) Schematic of 40 nm LSSO / 8 nm SSO / GSO (110) films with fixed layer thickness and different La (dopant) concentrations and (d) corresponding X-ray diffraction patterns; (e) and (f) s-SNOM images of two films grown with different La concentrations – $T_{La}$ = 1150°C and 1200°C, showing the tunability of the microstructure and resultant dielectric behaviors with chemical doping.

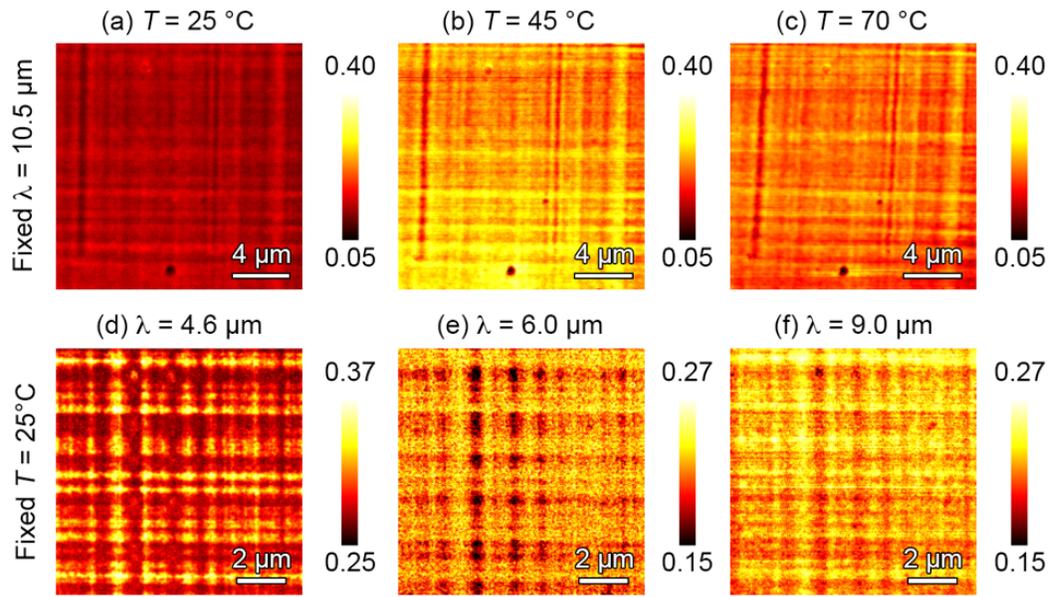

**Figure 5:** s-SNOM of 35 nm LSSO / 7 nm SSO / GSO (110) film ($T_{La}$ = 1150 °C) as a function of temperature, $T$, at fixed wavelength, $\lambda$ = 10.5 µm (a–c), and as a function of wavelength, $\lambda$, at fixed temperature, $T$ = 25 °C (b–d).

# Supplementary Materials for

# Self-Assembled Periodic Nanostructures Using Martensitic Phase Transformations

Abhinav Prakash[1,*,†], Tianqi Wang[1,†], Ashley Bucsek[2,3,†], Tristan K. Truttmann[1], Alireza Fali[4], Michele Cotrufo[5], Hwanhui Yun[1], Jong-Woo Kim[6], Philip J. Ryan[6], K. Andre Mkhoyan[1], Andrea Alù[5,7], Yohannes Abate[4], Richard D. James[2], and Bharat Jalan[1,*]

[1]Department of Chemical Engineering and Materials Science, University of Minnesota – Twin Cities, Minneapolis, MN 55455, USA.

[2]Department of Aerospace Engineering and Mechanics, University of Minnesota – Twin Cities, Minneapolis, MN 55455, USA.

[3]Department of Mechanical Engineering, University of Michigan, Ann Arbor, MI 48109, USA.

[4] Department of Physics and Astronomy, University of Georgia, Athens, GA 30602, USA.

[5]Photonics Initiative, Advanced Science Research Center, City University of New York, NY 10031, USA.

[6]Advanced Photon Source, Argonne National Laboratory, Argonne, IL 60439, USA.

[7]Physics Program, Graduate Center, City University of New York, New York, NY 10016, USA.

*Correspondence to: Abhinav Prakash (praka019@umn.edu), Bharat Jalan (bjalan@umn.edu).

†Equal contributions.

**This PDF file includes:**

    Supplementary Text
    Figs. S1 to S6

# Supplementary Text

## I. Characterization of periodic microstructure through synchrotron X-rays

Figure S1a shows temperature-dependent wide-angle X-ray diffraction scans for 40 nm La-doped SSO film grown on 8 nm SSO/GSO (110). The La-dopant cell temperature was kept at 1150 °C. Figure S1b shows the temperature-dependent rocking curve (H-scan) taken at the Bragg peak (right film peak) for the orthorhombic SSO phase showing a periodic modulation in the intensity across the center film peak suggesting periodic microstructure consistent with the results of SNOM (Figure 4e of the main text). Also shown is the two-dimensional HK-mesh scan for the orthorhombic phase at 180 K showing modulation in intensity along H- and K-directions. Periodicity of these intensity modulations reveals a periodicity of 40-50 nm at 180 K and 60-70 nm at 300 K, which is again consistent with the observation in the SNOM results of Fig. 4e in the main text. Additionally these results also reveal that high temperature tetragonal phase remain locked in within the orthorhombic phase below the phase transition temperature resulting in a finer microstructure (Fig. S1d).

## II. Surface morphology

Figure S2 shows the surface morphology of 30 nm $La_xSr_{1-x}SnO_3/t_{buffer}$ $SrSnO_3/GdScO_3$ (110) films as a function of buffer layer thicknesses ($t_{buffer}$) showing vertical stripes for thinner buffer layers followed by both vertical and horizontal stripes for thicker buffer layered films. No correlation exists between the height contrast in these AFM images and the SNOM contrast in the Figs. 2, 4 and 5 in the main text. To investigate these strips, detailed STEM studies were carried out. Figure S3 shows a representative low magnification STEM image (top panel) and a high-resolution HAADF STEM image (bottom panel) of a 30 nm $La_xSr_{1-x}SnO_3/108$ nm

SSO/GSO (110) film. Multiple steps consistent with those observed in AFM were observed. It is noted that these step edges do not coincide with the phase boundary.

**III. Comparison of s-SNOM Contrasts and Topographical changes**

To show that the observed IR near-field contrasts are attributable to changes in the dielectric environment and not to topographical changes along the out-of-plane axis, we compared spatial maps of the sample with s-SNOM amplitude measurements (Fig. S4). Comparing line scans at the same topographical location in both 2$^{nd}$ harmonic IR amplitude s-SNOM signal and topographical height, we see that while there appears to be small topographic variations (due to noise induced residual contaminants from the sample fabrication process), these do not correlate with the location of domains in the IR s-SNOM amplitude maps. This clearly demonstrates that the domains observed in the IR s-SNOM maps are due to existence of two kinds of materials with different dielectric functions, not to changes in the topography of the sample surface.

**IV. Strain Relaxation**

To investigate the strain relaxation in SSO films as a function of $t_{buffer}$, high resolution reciprocal space mapping (RSM) was performed on a representative film with the $t_{buffer}$ = 144 nm. Figure S5 shows a RSM scan along (103) planes indicating film remain completely coherent to the underlying substrate. This result in combination with the figure 4b of the main text confirms that strain relaxation in the film occurs via the martensitic phase transformation from the tetragonal-to-orthorhombic phase upto a thickness, $t_{buffer}$ = 144 nm.

**V. Effect of doping on phase transformation temperature**

When electronic carriers in a solid are added through doping, they may alter the Gibbs free energy landscape and therefore may change the phase transition temperature or allow for an access of different phase at a fixed temperature as a function of carrier density. Figure S6 show

T-dependent RSM scans of a 40 nm SSO film grown on 8 nm SSO/GSO (110) with and without La doping indicating that the addition of La-dopants (< 0.1 at.%) decrease the amount of orthorhombic phase at RT and change the transformation temperature (for tetragonal → orthorhombic) to a higher temperature in consistent with the results of Figure 4d in the main text.

## VI. Crystallographic Theory of Martensite Calculations

Martensitic phase transformations are diffusionless transformations between a high-symmetry crystallographic phase and a low-symmetry crystallographic phase. In this section, we will refer to the high-symmetry crystallographic phase as austenite, and we will refer to the low-symmetry crystallographic phase as martensite. The crystallographic theory of martensite (CTM), also called the phenomenological theory of martensite or the geometrically nonlinear theory of martensite, uses the crystallography of the two phases to predict all compatible interfaces and orientation relationships based on kinematic compatibility[1-6]. The theory uses the stretch component of the transformation that takes the austenite unit lattice to the martensite unit lattice. Because the martensite phase is less symmetric than the austenite phase, there are several different orientations of the martensite relative to the austenite. These are called crystallographic variants, or correspondence variants (CVs), and the number of CVs is equal to the number of rotations in the austenite Laue point group divided by the number of rotations in the martensite Laue point group.

The tetragonal-to-orthorhombic phase transformation has two martensite CVs of types

$$\boldsymbol{U_1} = \begin{bmatrix} \alpha & 0 & 0 \\ 0 & \beta & 0 \\ 0 & 0 & \gamma \end{bmatrix}, \boldsymbol{U_2} = \begin{bmatrix} \beta & 0 & 0 \\ 0 & \alpha & 0 \\ 0 & 0 & \gamma \end{bmatrix} \qquad (1)$$

where are the stretch ratios of the orthorhombic lattice parameters to the tetragonal lattice parameters, $\alpha = a^O/a^T$, $\beta = b^O/a^T$, and $\gamma = c^O/c^T$. Superscripts $O$ and $T$ refer to the orthorgonal and tetragonal phases, respectively.

In a bulk material, an interface between austenite and a single martensite CV is not usually possible (i.e., kinematically compatible), but an interface between austenite and twinned martensite is approximately compatible in an average, continuum sense. Let us consider if a compatible interface between austenite and twinned martensite can form in a bulk materials. We work with respect to the austenite unit cell, so the deformation on the austenite side of the interface is $I$, where $I$ is the identity tensor. The deformation on the martensite side of the interface is the average deformation of two martensite variants $U_i$ and $U_j$ that form twins. First, the twins must be compatible with each other at the twin plane. To form a compatible twin plane, the two martensite CVs must satisfy the compatibility condition

$$(QU_i - U_j) = a \otimes \hat{n} \qquad (2)$$

where $\hat{n}$ is the twin plane outward normal and $a$ is called the shear direction. Next, to form a compatible interface between austenite and the martensite twins defined in (2), we would need to satisfy the compatibility condition

$$\hat{\hat{Q}}(\lambda QU_i + (1-\lambda)U_j) - I = b \otimes \hat{m} \qquad (3)$$

where $\hat{m}$ is the outward normal of the austenite-twinned martensite interface and $b$ is called the shape strain direction. The methodology for calculating the solutions to (2) and (3) can be found in[2-4].

In the case of a thin film, the compatibility requirements are less stringent, so we can consider a possible compatible interface between austenite and a single martensite CV, even in the case where such an interface is not possible in a bulk material.[2,7] The compatible interface is reduced from a plane to a line that is perpendicular to the outward normal of the substrate-film interface (i.e., perpendicular to the out-of-plane growth direction), and the condition for compatibility becomes

$$(QU_i - I)\hat{e} = 0, \quad \hat{e} \cdot \hat{e}_3 = 0 \qquad (4)$$

where the out-of-plane direction is $\hat{e}_3$ and the compatible line interface is $\hat{e}$.

The following methodology for solving (4) is published in reference[7] and reprinted in Chapter 10 of [2]. It is summarized here for interested readers.

1. Calculate

$$A = U_i^T U_i - I. \qquad (5)$$

2. Equation (4) has a solution if and only if

$$\hat{e}_3 \cdot cof A \hat{e}_3 \leq 0. \qquad (6)$$

3. To find $\hat{e}_3$, choose any mutually perpendicular unit vectors $\hat{e}_1$ and $\hat{e}_2$ in the film plane so that $\{\hat{e}_1, \hat{e}_2, \hat{e}_3\}$ form an orthonormal basis. The solution to (4) is given by

$$\hat{e} = \alpha \hat{e}_1 + \beta \hat{e}_2 \qquad (7)$$

where $\alpha, \beta$ simultaneously satisfy the equations

$$\alpha^2 + \beta^2 = 1, \qquad (8)$$

$$\alpha^2 \hat{e}_1 \cdot A\hat{e}_1 + 2\alpha\beta \hat{e}_1 \cdot A\hat{e}_2 + \beta^2 \hat{e}_2 \cdot A\hat{e}_2 = 0. \qquad (9)$$

For each solution, the necessary rotation $Q$ on $U_i$ from (4) can be found by

$$\hat{f} = \frac{U_i \hat{e} \times \hat{e}}{|U_i \hat{e} \times \hat{e}|} \qquad (10)$$

$$\theta = \text{acos}\,(\hat{e} \cdot U_i \hat{e}) \qquad (11)$$

$$Q = R_{\hat{e}}(\varphi) Q_{\hat{f}}(\theta) \qquad (12)$$

where $Q_{\hat{f}}(\theta)$ is a rotation $\theta$ about $\hat{f}$, and $R_{\hat{e}}(\varphi)$ is any rotation $\hat{e}$. Thus, there is a set of possible rotations on $U_i$ for which $\hat{e}$ is a solution.

The transformation strain, or strain relief that would be produced in the case of an epitaxial film, can be calculated using

$$\boldsymbol{\varepsilon} = \begin{bmatrix} \varepsilon_{11} & \varepsilon_{12} & \varepsilon_{13} \\ \varepsilon_{12} & \varepsilon_{22} & \varepsilon_{23} \\ \varepsilon_{13} & \varepsilon_{23} & \varepsilon_{33} \end{bmatrix} = \frac{1}{2}(\boldsymbol{U}_i^T \cdot \boldsymbol{U}_i - \boldsymbol{I}). \qquad (13)$$

Special attention should be made to the coordinate system, as $\boldsymbol{U}_i$ is defined relative to the austenite phase unit cell. If, for example, the austenite phase is rotated as in the case of a (110) film, then the proper coordinate transformations must be applied. This will affect the results of both (4) and (13).

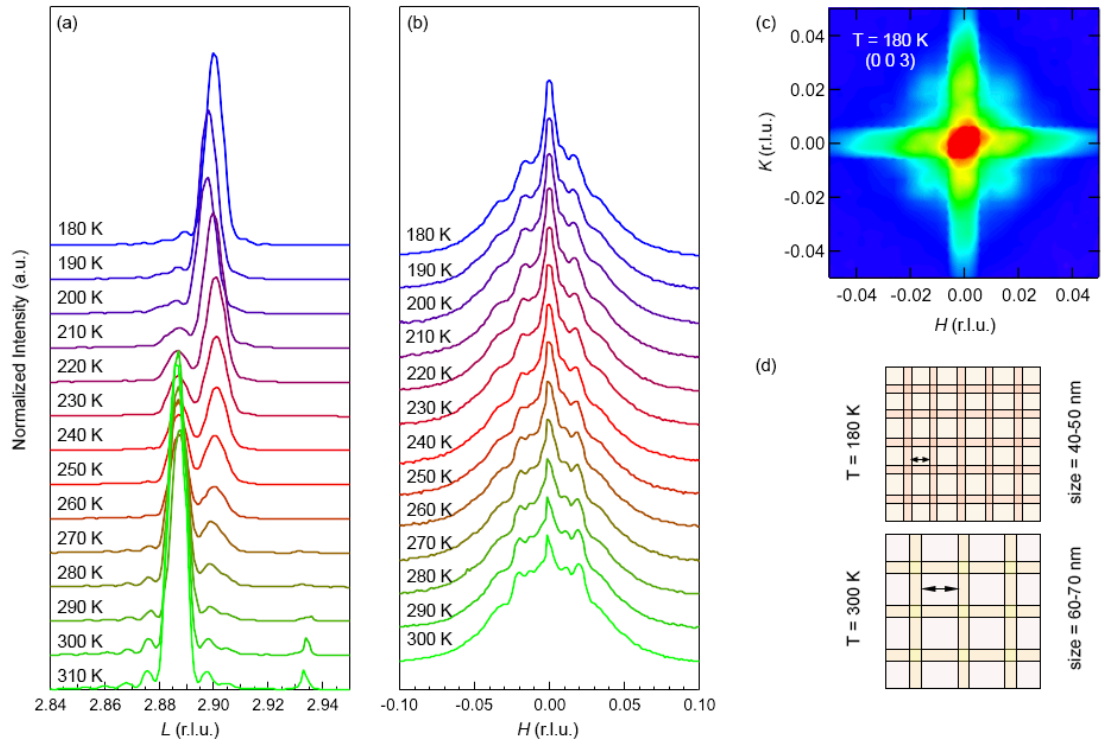

**Figure S1**: (a) Wide-angle X-ray diffraction scans as a function of temperature for 40 nm $La_xSr_{1-x}SnO_3$/8 nm SSO/GSO (110) grown with $T_{La}$ = 1150 °C, (b) Rocking curve (***H***-scan) taken at the Bragg condition (right film peak) for the orthorhombic (low-temperature) $SrSnO_3$ phase, (c) ***HK*** mesh scan for the orthorhombic phase at 180 K, and (d) Schematic representation of the structure expected based on the rocking curve scans. The domain size was estimated based on the reciprocal spacing between the intensity modulations observed in the rocking curve shown in (b).

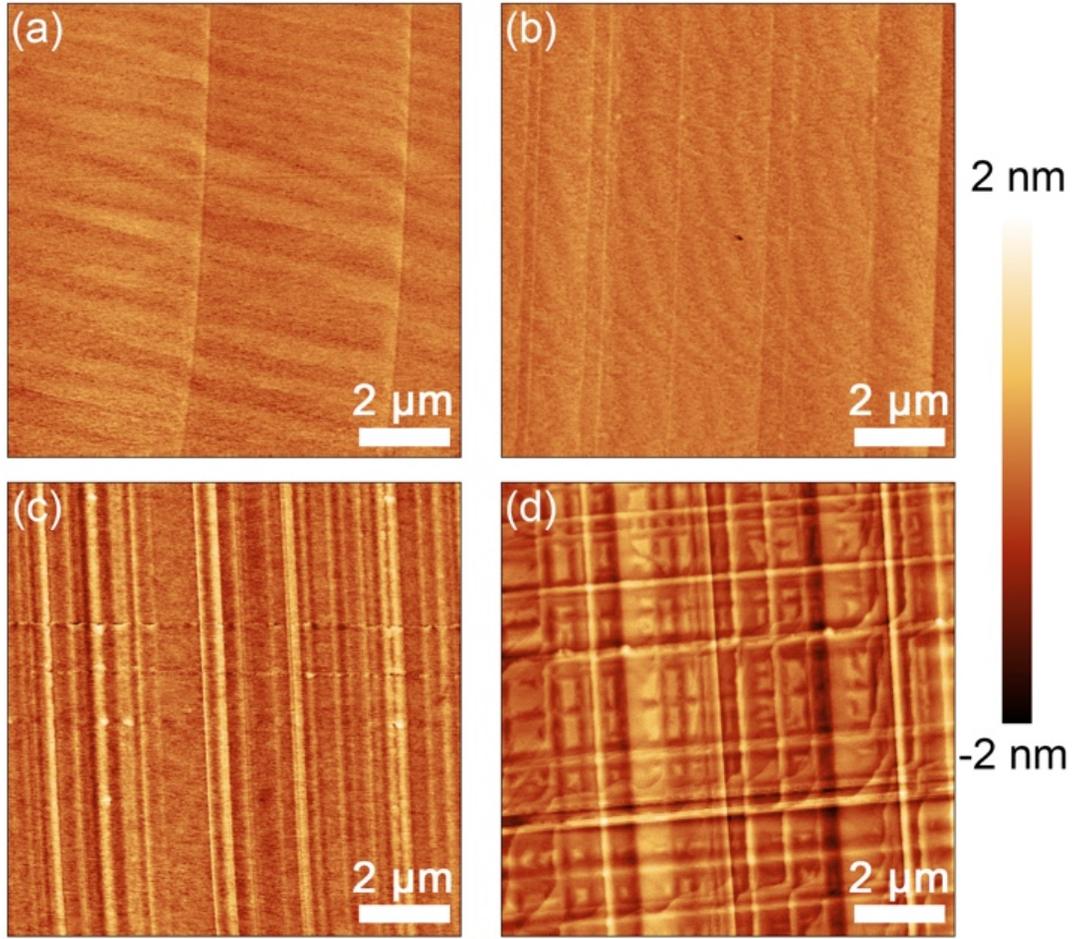

**Figure S2:** Atomic force microscopy (AMF) images of thin films of 30 nm $La_xSr_{1-x}SnO_3/t_{buffer}$ $SrSnO_3/GdScO_3$ (110) with $t_{buffer}$ = (a) 2 nm, (b) 39 nm, (c) 72 nm, and (d) 108 nm. The root mean square roughnesses were 2.0 Å, 1.6 Å, 4.1 Å, and 4.5 Å. With increasing thickness, it is observed that the density lines increase – first in one (vertical) direction and subsequent (horizontal) lines start to appear in the orthogonal direction. Figure S2 is used to identify whether these lines are atomic steps due to step-flow growth or steps due to the presence of phase boundary shown in figure 1(g).

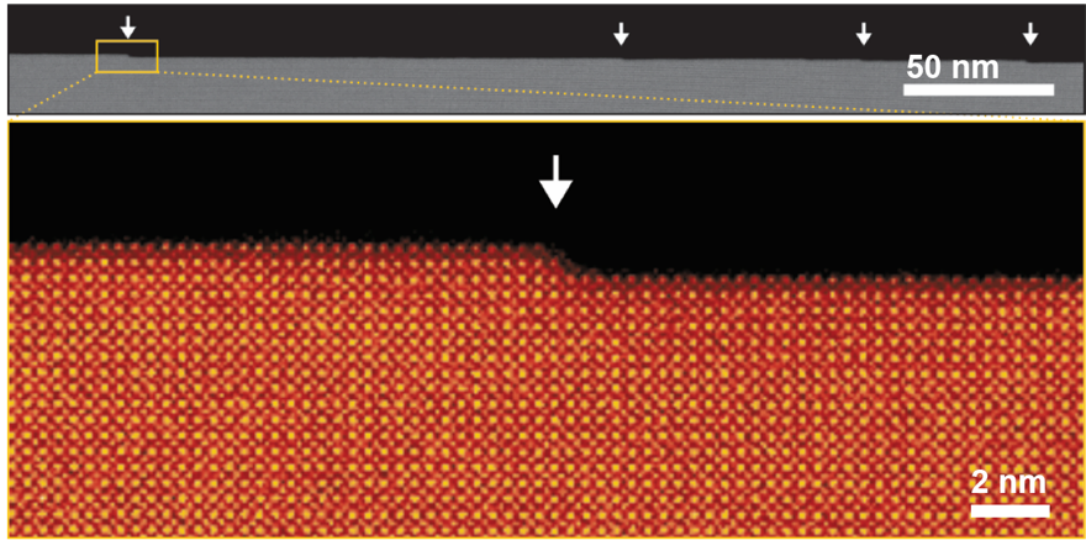

**Figure S3:** Top panel shows a low magnification transmission electron microscopy image of the 30 nm La$_x$Sr$_{1-x}$SnO$_3$/108 nm SrSnO$_3$/GdScO$_3$ (110) film showing the presence of atomic steps on the surface. Bottom panel shows a high-resolution scanning transmission electron microscopy image of one of these steps. These steps did not coincide with the phase boundary shown in the main text figure 1(g) suggesting that atomic force microscopy could be misleading in identifying the two phases and characterization tools such as scanning near-field optical microscopy becomes indispensable for phase classification. The step height was approximately 2-unit cells.

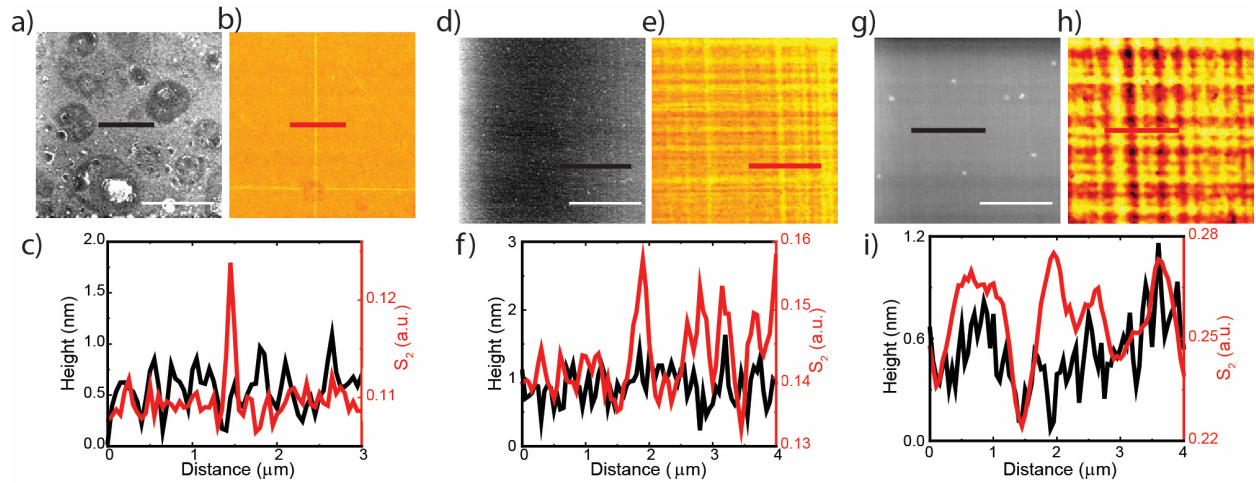

**Figure S4:** Topography (black/white, (a), (d) and (g)), and near-field 2nd Harmonic IR amplitude (yellow/red, (b), (e) and (h)). Correlated line profiles (lower panel, (c), (f) and (i). Figures (b), (e) and (h) are the same figures in the main text as Figure 2 (d), (e) and (f) respectively. Scale bar is 4 μm.

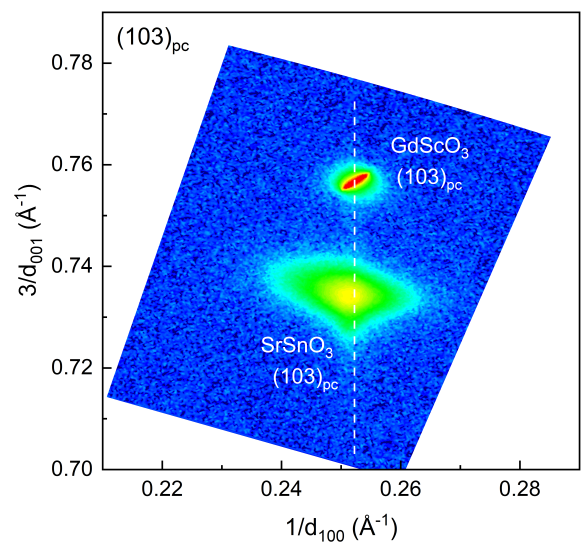

**Figure S5**: Reciprocal space map for a 30 nm $La_xSr_{1-x}SnO_3$/144 nm $SrSnO_3$/$GdSnO_3$ (110) films around the $(103)_{pc}$ Bragg reflection of the substrate suggesting the film was predominately strained.

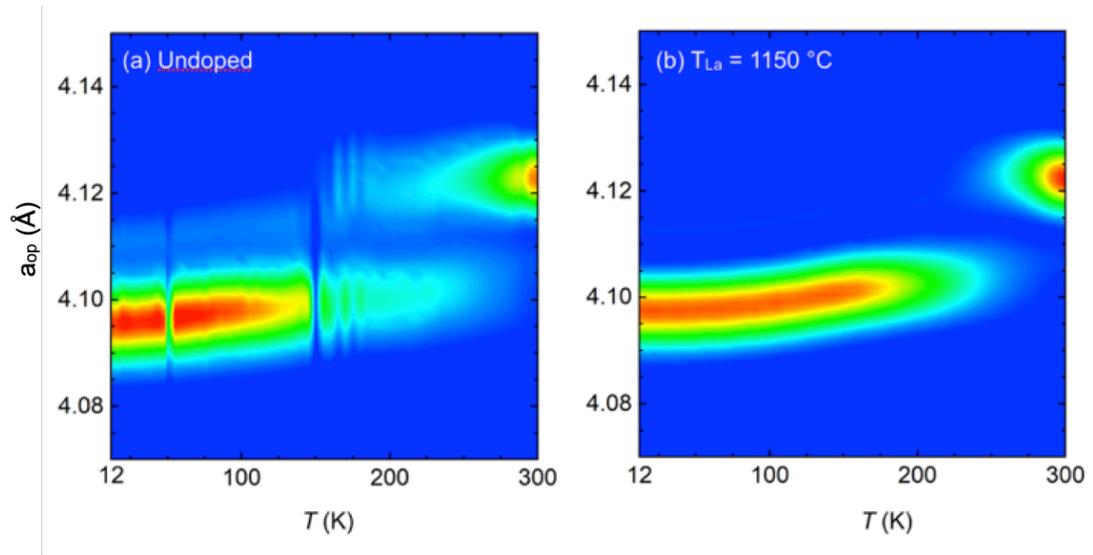

**Figure S6:** Temperature-dependent out-of-plane lattice parameters a 40 nm SSO film grown on 8 nm SSO/GSO (110) (a) without La and (b) with La doping ($T_{La}$ = 1150 °C)